\documentclass[twocolumn,pre,aps,nofootinbib,superscriptaddress,showpacs,preprintnumbers]{revtex4}

\usepackage{amssymb,amssymb,epsfig,bm}

\begin{document}

\title{Vorticity production and survival in viscous and magnetized cosmologies}

\author{F. Dosopoulou}
\affiliation{Section of Astrophysics, Astronomy and Mechanics, Department of Physics, Aristotle University of Thessaloniki, Thessaloniki 54124, Greece}
\affiliation{Department of Physics, University of Crete, Heraklion 71003, Greece}

\author{F. Del Sordo}
\affiliation{NORDITA, AlbaNova University Center, Roslagstullsbacken 23, SE-10691 Stockholm, Sweden}
\affiliation{Department of Astronomy, Stockholm University, SE-10691 Stockholm, Sweden}

\author{C.G. Tsagas}
\affiliation{Section of Astrophysics, Astronomy and Mechanics, Department of Physics, Aristotle University of Thessaloniki, Thessaloniki 54124, Greece}
\affiliation{NORDITA, AlbaNova University Center, Roslagstullsbacken 23, SE-10691 Stockholm, Sweden}

\author{A. Brandenburg}
\affiliation{NORDITA, AlbaNova University Center, Roslagstullsbacken 23, SE-10691 Stockholm, Sweden}
\affiliation{Department of Astronomy, Stockholm University, SE-10691 Stockholm, Sweden}

\begin{abstract}
We study the role of viscosity and the effects of a magnetic field on a rotating, self-gravitating fluid, using Newtonian theory and adopting the ideal magnetohydrodynamic approximation. Our results confirm that viscosity can generate vorticity in inhomogeneous environments, while the magnetic tension can produce vorticity even in the absence of fluid pressure and density gradients. Linearizing our equations around an Einstein-de Sitter cosmology, we find that viscosity adds to the diluting effect of the universal expansion.
Typically, however, the dissipative viscous effects are confined to relatively small scales. We also identify the characteristic length bellow which the viscous dissipation is strong and beyond which viscosity is essentially negligible. In contrast, magnetism seems to favor cosmic rotation. The magnetic presence is found to slow down the standard decay-rate of linear vortices, thus leading to universes with more residual rotation than generally anticipated.
\end{abstract}


\maketitle

\section{Introduction}
Most, if not all, astrophysical systems rotate and it is conceivable that vorticity can play an important role in a number of situations. So far, however, the majority of the related studies has been confined to relatively small-scale systems. Although the question of universal rotation and how large this may be has been posed more than once (see, e.g.,~\cite{BJS,KO,KHB,GSFB,JBEGH,SC} and references therein), in cosmology the role of vorticity is usually bypassed. One of the reasons for doing so is that typical inflationary scenarios are irrotational, since the vacuum fluctuations do not generate vorticity~\cite{LL}. Another, is that linear rotational perturbations are known to decay as the inverse square of the cosmological scale factor, which implies strong dilution for any vortex-like distortions that could happen to exist. Dissipative effects during the subsequent radiation era, like Silk damping and neutrino free-streaming, may also have a negative impact on large-scale rotation, though (to the best of our knowledge) studies of this nature are not yet available in the literature.

In cosmology, where the terms vorticity and rotation are used interchangeably, vorticity is usually generated from nonlinear
effects. This is one additional reason for not taking vorticity into account in many cosmological perturbation studies. Inhomogeneous non-barotropic (e.g., non-adiabatic) media, for example, can generate rotation at the second perturbative level~\cite{CMM,CM}, and so can fluid viscosity~\cite{MB,D-SB,Fetal}. Nevertheless, resorting to nonlinear effects is not always necessary, since there are agents that can induce rotation still at the linear level. For instance, a coherent magnetic field (`$B$-field') can act as a source of cosmic vorticity~\cite{W,TM}. The opposite is also theoretically possible. Thus, following~\cite{H}, a number of authors have explored the likelihood that rotational distortions can lead to cosmic magnetogenesis (see, e.g.,~\cite{Metal,MMNR,KMST,GR} and references therein). In addition, both the fluid viscosity and the $B$-field can affect the evolution of rotational perturbations and thus change the standard picture, which usually assumes non-magnetized, perfect media. Therefore, in principle at least, the effects of the aforementioned sources could lead to more residual vorticity than previously anticipated.

The present paper uses Newtonian gravity to study the effects of viscosity and those of a magnetic field on rotation. We adopt the covariant, Lagrangian approach to fluid dynamics, which has been proved helpful in several (mainly relativistic) cosmological studies (see~\cite{E1} for an introduction to the Newtonian version of the formalism). First, we look into the effects of fluid viscosity and then consider those of a coherent magnetic field. Our results, confirm those of earlier studies claiming that both of the aforementioned agents can act as sources of cosmic rotation. We show, in particular, that viscosity can only generate vorticity in media with an inhomogeneous density distribution. This is not the case for the $B$-field, which does not (in principle at least) require the presence of density perturbations. Assuming the Newtonian analogue of the Einstein-de Sitter universe as our background model, we find that viscosity can only generate vorticity at second order. Magnetism, on the other hand, can do the same at the linear perturbative level. Moreover, the irreducible decomposition of variables and equations that the covariant formalism introduces, allows us to identify the agents leading to vorticity generation, as well as the way they do so. Viscosity, for example, sources rotational distortions via the coupling of the shear with certain types of density perturbations. Alternatively, the magnetic field acts primarily through the tension component of the Lorentz force.

Our next step is to investigate how viscous and/or magnetized media can affect the linear evolution of rotational perturbations. In the first case, we find that viscosity always leads to vorticity dissipation, though the effects are generally scale dependent. We identify, in particular, a characteristic length beyond which viscosity is essentially negligible and below which its role becomes progressively more important. In particular, well inside this ``viscous scale'' vorticity dissipates very quickly, while far beyond that scale the fluid kinematics are dominated by the background Hubble expansion. There, one recovers the standard inverse-square decay law for rotational perturbations. The size of the aforementioned viscous scale depends on the specifics of the cosmic fluid, such as the average velocity and the mean free-path of the species involved. When dealing with non-relativistic matter, however, the typical domain affected by viscosity is very small (in cosmological terms).

In contrast to viscosity, the presence of magnetic fields seems to have a favorable effect on linear rotational perturbations. In addition to generating vorticity at the linear order, the $B$-field slows down the standard decay rate of rotation in perturbed Friedmann-Robertson-Walker (FRW) universes. As mentioned earlier, vorticity typically depletes as $a^{-2}$, with $a$ representing the cosmological scale factor. When the linear magnetic effects are accounted for, however, the vorticity decay-rate changes to $a^{-3/2}$. The implication of this effect, which comes entirely from the magnetic tension, is that magnetized universes, can have more residual vorticity than their non-magnetic counterparts.
An analogous, conclusion was also reached indirectly, through the relativistic analysis of linear density inhomogeneities in magnetized cosmologies~\cite{TM}. To the best of our knowledge, this is the first direct indication that magnetism ``favors'' both the generation and the survival of cosmic rotation.

\section{Newtonian self-gravitating media}
\subsection{Newtonian kinematics}
Consider a self-gravitating Newtonian medium and a velocity field $v_a$, tangent to the flow lines of the fluid. The local kinematics is analyzed by splitting the gradient $\partial_bv_a$ into its irreducible components according to~\cite{E1}
\begin{equation}
\partial_bv_a= {1\over3}\,\Theta\delta_{ab}+ \sigma_{ab} +\epsilon_{abc}\omega^c\,,  \label{pbva}
\end{equation}
where $\Theta=\partial^av_a$ is the volume scalar, $\sigma_{ab}=\partial_{\langle b}v_{a\rangle}$ represents the shear tensor and $\omega_a=-{\rm curl}v_a/2$ defines the vorticity vector.\footnote{Round brackets denote symmetrization, square antisymmetrization and angled ones indicate the symmetric and trace-free part of second-rank tensors (e.g.~$\sigma_{ab}=\partial_{\langle b}v_{a\rangle}= \partial_{(b}v_{a)}-(\Theta/3)\delta_{ab}$).} Also, $\delta_{ab}$ is the Kronecker symbol and $\varepsilon_{abc}$ is the Levi-Civita tensor. By construction, the volume scalar determines the expansion or contraction of a fluid element (when positive or negative, respectively) and defines a representative length scale ($a$) by means of $\dot{a}/a=\Theta/3$. The shear, on the other hand, describes changes in the element's shape (under constant volume) and $\omega_a$ defines the local rotational axis.

The above defined variables obey a set of three evolution and three constraint equations that fully determine the fluid kinematics. In particular, the propagation formulae are~\cite{E1,ST}
\begin{equation}
\dot{\Theta}= -{1\over3}\,\Theta^2- {1\over2}\,\kappa\rho+ \partial^aA_a- 2\left(\sigma^2-\omega^2\right)\,,  \label{Ray}
\end{equation}
\begin{equation}
\dot{\sigma}_{ab}= -{2\over3}\,\Theta\sigma_{ab}- E_{ab}+ \partial_{\langle b}A_{a\rangle}- \sigma_{c\langle a}\sigma^c{}_{b\rangle}- \omega_{\langle a}\omega_{b\rangle} \label{dotsigma}
\end{equation}
and
\begin{equation}
\dot{\omega}_a= -{2\over3}\,\Theta\omega_a- {1\over2}\,{\rm curl}A_a+ \sigma_{ab}\omega^b\,,  \label{dotomega}
\end{equation}
where overdots denote convective derivatives (e.g.~$\dot{\Theta}=
\partial_t\Theta+v^b\partial_b\Theta$). Here, $\rho$ is the density of the matter (with $\kappa=8\pi G$ being the rescaled gravitational constant), while $2\sigma^2=\sigma_{ab}\sigma^{ab}$ and $2\omega^2=\omega_{ab}\omega^{ab}$ define the magnitudes of the shear and the vorticity, respectively. We also note the vector $A_a=\dot{v}_a+\partial_a\Phi$, which is the sum of the inertial and gravitational acceleration ($\Phi$ is the gravitational potential) and vanishes in the absence of non-inertial or non-gravitational forces.\footnote{Putting the inertial and the gravitational acceleration together means that there is no longer an explicit need for Poisson's equation. The latter, however, has been employed when deriving the set of (\ref{Ray})--(\ref{dotomega}) and takes the
form $\partial^2\Phi=\kappa\rho/2$ in our notation, with $\partial^2=\partial^a\partial_a$ representing the Laplacian operator.} Finally, the tensor $E_{ab}=\partial_{\langle a}\partial_{b\rangle}\Phi$ describes the tidal part of the gravitational field.

Expressions (\ref{Ray})--(\ref{dotomega}) are supplemented by an equal number of constraint equations. These relate the gradients of the key kinematic variables and are given by
\begin{equation}
\partial^a\omega_a= 0\,, \quad {\rm curl}\sigma_{ab}= -\partial_{\langle b}\omega_{a\rangle}  \label{con1}
\end{equation}
and
\begin{equation}
\partial^b\sigma_{ab}= {2\over3}\,\partial_a\Theta+ {\rm curl}\omega_a\,,  \label{con2}
\end{equation}
with
${\rm curl}\sigma_{ab}=\varepsilon_{cd\langle a}\partial^c\sigma^d{}_{b\rangle}$by definition~\cite{E1,ST}. The sets (\ref{Ray})--(\ref{dotomega}) and (\ref{con1})--(\ref{con2}) monitor the kinematics between two neighboring flow lines. To close the system one also needs the continuity equation, which reflects mass conservation and in our case is
\begin{equation}
\dot{\rho}= -\Theta\rho\,,  \label{cont}
\end{equation}
together with the momentum-conservation formula. The exact form of the latter depends on the type (i.e.~on the equation of state) of the matter and on the presence of additional sources.

\subsection{Newtonian cosmologies}
Newtonian cosmology is expected to provide a good approximation of the post-recombination universe and on scales well inside the Hubble radius, where one can safely ignore the role of spacetime curvature. In Newtonian gravity, the relativistic FRW universes are represented by models where the fundamental observers move along parabolic, elliptical and hyperbolic trajectories. These flow lines correspond to Euclidean (flat), spherical (closed) and hyperbolic (open) spatial geometry respectively. Given the homogeneity and isotropy of the Friedmann universes, the only nonzero variables are scalars that depend solely on time. Vectors, tensors and spatial gradients vanish identically. Thus, the Newtonian analogue of a dust-dominated, flat FRW cosmology (i.e.~of the Einstein-de Sitter universe) is monitored by the set
\begin{equation}
H^2= {1\over3}\,\kappa\rho\,, \quad \dot{H}= -H^2- {1\over6}\kappa\rho  \label{N1}
\end{equation}
and
\begin{equation}
\dot{\rho}= -3H\rho\,.  \label{N2}
\end{equation}
These relations are respectively known as the Friedmann, the Raychaudhuri and the continuity equations. Note that $H=\dot{a}/a$ is the Hubble parameter, with $a$ representing the cosmological scale factor. This implies that $H=\Theta/3$ in all FRW cosmologies. It is then relatively straightforward to show that (\ref{N1}b) and (\ref{N2}) are the Einstein-de Sitter limits of (\ref{Ray}) and (\ref{cont}) respectively. Friedmann's equation, on the other hand, is simply an integral of (\ref{N1}b).

The system of (\ref{N1}) and (\ref{N2}) can be solved analytically with respect to time, leading to
\begin{equation}
a\propto t^{2/3}\,, \quad H= {2\over3t} \quad {\rm and} \quad \kappa\rho= {4\over3t^2}\,,  \label{EdS}
\end{equation}
which offer an alternative description of the Einstein-de Sitter universe. In the following sections, we will consider linear rotational perturbations around such a background, and study their evolution in the presence of fluid viscosity or a magnetic field.

\section{Viscous rotation}
\subsection{Viscosity induced vorticity}
The rotational evolution of a self-gravitating Newtonian fluid is governed by Eq.~(\ref{dotomega}), which determines the twist between two neighboring flow lines. Note the first and the last terms on the right-hand side of that relation. These describe effects triggered by the relative motion of the fluid particles, analogous to those attributed to the Coriolis force~\cite{D-SB}. The second term, on the other hand, reflects the presence of non-inertial and non-gravitational forces (e.g.~those triggered by pressure gradients, the presence of electromagnetic fields, etc) and is determined by the associated Navier-Stokes equation. In the case of a (non-magnetized) viscous fluid, the latter reads~\cite{E1}
\begin{equation}
\rho A_a= -\partial_ap- \partial^b\pi_{ab}\,,  \label{vNS}
\end{equation}
where $p$ is the isotropic pressure and $\pi_{ab}$ is the anisotropic pressure of the matter (with $\pi_{ab}=\pi_{ba}$ and $\pi_a{}^a=0$). Taking the ``curl'' of the above and then substituting into expression (\ref{dotomega}) provides an evolution formula that monitors the rotational behavior of a Newtonian viscous medium.

Fluid viscosity can be expressed in terms of kinematic (shear) viscosity via the phenomenological relation
\begin{equation}
\pi_{ab}= -2\nu\rho\sigma_{ab}\,,  \label{nu1}
\end{equation}
with $\nu$ representing the kinematic viscosity coefficient and $\nu>0$~\cite{D-SB}. The latter immediately implies that $\nu=\pi/(2\rho\sigma)$, where $2\pi^2=\pi_{ab}\pi^{ab}$ defines the magnitude of the anisotropic pressure tensor. In addition, when the viscosity coefficient is constant, the divergence of Eq.~(\ref{nu1}) ensures that
\begin{equation}
\partial^b\pi_{ab}= -2\nu\sigma_{ab}\partial^b\rho- {4\over3}\,\nu\rho\partial_a\Theta- 2\nu\rho{\rm curl}\omega_a\,,  \label{scon}
\end{equation}
where we have used the so-called shear constraint (see Eq.~(\ref{con2}) earlier).

Inserting the auxiliary relation (\ref{scon}) into the right-hand side of (\ref{vNS}), taking the ``curl'' of the resulting expression, using the constraint (\ref{con1}a) and then substituting the outcome into Eq.~(\ref{dotomega}), the latter gives
\begin{eqnarray}
\dot{\omega}_a= &\!-\!& {2\over3}\,\Theta\omega_a-
{1\over2\rho}\,\varepsilon_{abc}\partial^b\left(\ln\rho\right)
\partial^cp \nonumber\\ &\!-\!& \nu\varepsilon_{abc}\partial^b
\left[\sigma^{cd}\partial_d\left(\ln\rho\right)\right]
+\nu\partial^2\omega_a \nonumber\\ &\!+\!& \sigma_{ab}\omega^b\,.
\label{vdotomega}
\end{eqnarray}
According to the above, in the absence of viscosity, the only way of generating vorticity is through the baroclinic (the second) term on the right-hand side of (\ref{vdotomega}). The latter vanishes when $p=p(\rho)$, namely in the case of a barotropic medium. The viscous nature of the fluid, on the other hand, can act as a source of rotation by exploiting the coupling between the density and the shear gradients. Analogous conclusions have also been reached in~\cite{D-SB}. In fact, after removing the baroclinic term and taking into account the difference in the definition of the vorticity vector, one can show that expression (\ref{vdotomega}) is identical to Eq.~(1) in~\cite{D-SB}.

Let us consider the viscous source-term in the right-hand side of (\ref{vdotomega}), the significance of which means that it deserves further attention. Recalling that ${\rm curl}\sigma_{ab}= \varepsilon_{cd\langle a}\partial^c\sigma^d{}_{b\rangle}$ and using constraint (\ref{con1}b), a straightforward calculation leads to the expression (see also Appendix~\ref{A1})
\begin{eqnarray}
\varepsilon_{abc}\partial^b
\left[\sigma^{cd}\partial_d\left(\ln\rho\right)\right]=
&\!-\!& 2\Delta^b\partial_{\langle b}\omega_{a\rangle}- \Delta^b\varepsilon_{bcd}\partial^c\sigma^d{}_a \nonumber\\
&\!+\!& \varepsilon_{abc}\Sigma^{bd}\sigma^c{}_d\,,
\label{vsource}
\end{eqnarray}
which decomposes the source term into its irreducible components. According to the above, the source term is not entirely rotation-free, implying that only a part of it can actually generate vorticity. Also, $\Delta_a=\partial_a(\ln\rho)$ and
$\Sigma_{ab}=\partial_{\langle b}\partial_{a\rangle}(\ln\rho)$ by
definition. Following~\cite{E2}, the former describes the density contrast between two neighboring fluid flow lines, while the latter monitors changes (under constant volume) in the shape of an inhomogeneous density distribution (e.g.~from spherical symmetry to that of an ellipsoidal, etc). Note that, in contrast to \cite{E2}, here both $\Delta_a$ and $\Sigma_{ab}$ are not dimensionless.

\subsection{Viscous rotating universes}
Linearized around the Newtonian analogue of the Einstein-de Sitter
universe, where matter is in the form of pressureless matter (dust), expression (\ref{vdotomega}) reduces to\footnote{When linearizing, the only zero-order variables are the background ones. In an Einstein-de Sitter universe, these are the volume scalar (which coincides with the Hubble parameter -- i.e.~$H=\Theta/3$), the density and the isotropic pressure of the fluid. All these have only temporal dependence. Physical quantities that vanish in the background, but have nonzero perturbed value, like the vorticity, the viscous pressure, the shear and all the spatial gradients, are treated as first-order (linear) perturbations. Note that differentiating perturbations does not affect their perturbative order. Finally, products between linear quantities are of higher-order and are therefore neglected.}
\begin{equation}
\dot{\omega}_a= -2H\omega_a+ \nu\partial^2\omega_a\,, \label{EdSvdotomega}
\end{equation}
since $\Theta/3=H$ in the background. The above implies that, within the limits of the linear approximation, viscosity can no longer generate vorticity, though it has an effect on pre-existing rotation. According to (\ref{EdSvdotomega}), in the absence of viscous pressure, the vorticity evolution is determined by the background expansion only. In that case, we recover the standard decay-rate of $\omega_a\propto a^{-2}\propto t^{-4/3}$, since $a\propto t^{2/3}$ in an Einstein-de Sitter universe. The introduction of a viscous fluid, however, will generally change the standard picture. It is also worth noting that expression (\ref{EdSvdotomega}) looks like the ``heat-transfer equation'', with an extra term due to the universal expansion and the viscosity coefficient ($\nu$) playing the role of the thermal diffusivity.

\begin{figure}[tbp]
\begin{center}
\includegraphics[width=\columnwidth]{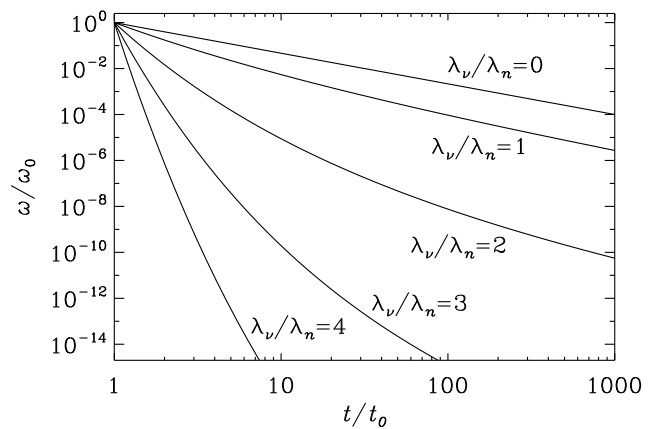}\quad
\end{center}
\caption{Evolution of rotational modes, with respect to time, on different scales; see Eq.~(\ref{vomega}). Sufficiently small scales are strongly affected by viscosity and there vorticity dissipates quickly. On larger wavelengths, however, rotational distortions can survive for long. There, at late times, vorticity is only affected by the background expansion. Note that the $\lambda_{\nu}/\lambda_n=0$ line corresponds to the standard $\omega\propto t^{-4/3}$ evolution law, while the
$\lambda_{\nu}/\lambda_n=1$ curve defines the viscous scale.} \label{fig1}
\end{figure}

To proceed, we introduce the harmonic splitting $\omega_a= \omega_{(n)}\mathcal{Q}_a^{(n)}$, with $\partial_a\omega_{(n)}=0= \dot{\mathcal{Q}}_a^{(n)}= \partial^a\mathcal{Q}_a^{(n)}$ and $\partial^2\mathcal{Q}_a^{(n)}=-(n/a)^2\mathcal{Q}_a^{(n)}$. Then, Eq.~(\ref{EdSvdotomega}) recasts into
\begin{equation}
\dot{\omega}_{(n)}= -2H\omega_{(n)}- \nu\left({n\over a}\right)^2\omega_{(n)}\,, \label{EdSnvdotomega1}
\end{equation}
where $n\geq0$ is the comoving wavenumber of the vorticity mode. Hence, qualitatively speaking, viscosity enhances the diluting role of the expansion and leads to less residual vorticity. The viscous effect, however, is scale dependent; see also Eqs.~(\ref{EdSnvdotomega3}) and (\ref{lambdanu}). Finally, recalling that $a\propto t^{2/3}$ and $H=2/3t$ in the background -- see~(\ref{EdS}), we arrive at
\begin{equation}
\dot{\omega}_{(n)}= -{4\over3t}\,\omega_{(n)}- \nu\left({n\over a_0}\right)^2\left({t_0\over t}\right)^{4/3}\omega_{(n)}\,, \label{EdSnvdotomega2}
\end{equation}
with the zero suffix indicating a given time during the evolution. Clearly, at sufficiently late times the last term on the right-hand side of the above becomes negligible and the viscosity effects fade away. More specifically, assuming that $\omega(t=t_0)=\omega_0$, the solution of (\ref{EdSnvdotomega2}) reads
\begin{eqnarray}
\frac{\omega}{\omega_0}=
\exp\left\{-4\left({\lambda_{\nu}\over\lambda_n}\right)_0^2
\left[1-\left({t_0\over t}\right)^{1/3}\right]\right\}
\left({t_0\over t}\right)^{4/3}\!,\quad
\label{vomega}
\end{eqnarray}
where $\lambda_{\nu}=\sqrt{\nu\lambda_H/2c}$ is a characteristic length that determines the domain of viscous influence; see Eq.~(\ref{lambdanu}) below. Also $\lambda_H=c/H$ (with $c$ being the speed of light) represents the Hubble horizon and $\lambda_n=a/n$ is the physical scale of the perturbation. In solution (\ref{vomega}) we see again that at late times the vorticity evolution is determined by the background expansion (see Fig.~\ref{fig1}). In particular, when $t\gg t_0$, the above expression is approximated by
\begin{equation}
\omega= \omega_0 \exp\left[-4\left({\lambda_{\nu}\over\lambda_n}\right)_0^2\right] \left({t_0\over t}\right)^{4/3}\,, \label{ltvomega}
\end{equation}
ensuring that $\omega\propto t^{-4/3}$. On the other hand, at sufficiently early times (i.e.~for $t\ll t_0$), the rotational behavior of the fluid is mainly dictated by the (viscous) exponential term of (\ref{vomega}). The overall effect, however, also depends on the scale of the perturbation.

\subsection{The viscous scale}
The scale dependence of the dissipative viscous effects is a key feature of solution (\ref{vomega}), because it determines their domain of influence and strength. As one can immediately see in Fig.~\ref{fig1}, the smaller the scale of the mode the stronger the viscous dissipation. Further understanding can be achieved through a simple qualitative analysis, without even solving Eq.~(\ref{EdSnvdotomega2}). To be precise, recalling that $\lambda_H=c/H$ and $\lambda_n=a/n$, expression (\ref{EdSnvdotomega1}) recasts as
\begin{equation}
\dot{\omega}_{(n)}= -2H\left(1+ {\nu\lambda_H\over2c\lambda_n^2}\right)\omega_{(n)}\,, \label{EdSnvdotomega3}
\end{equation}
Accordingly, the viscous effects are important when $\nu\lambda_H/2c\lambda_n^2\gg1$, which holds for modes with wavelength
\begin{equation}
\lambda_n\ll \lambda_{\nu}= \sqrt{\nu\lambda_H\over2c}\,.  \label{lambdanu}
\end{equation}
The right-hand side of the above defines a characteristic length, the ``viscous scale'' ($\lambda_{\nu}$), beyond which viscosity is essentially negligible and below which it becomes progressively more important.\footnote{On very small scales, we can no longer ignore the nonlinear effects and the linear approximation breaks down.} Using simple kinetic-theory arguments, one finds that $\nu\sim v\lambda_f$, where $v$ is the random velocity of the particles involved and $\lambda_f$ their mean free-path (see~\cite{BS} for further discussion and details). Hence, $\lambda_{\nu}\ll\lambda_H$ for non-relativistic species, which implies that the typical domain where the viscous effects can play a significant role is rather small. Furthermore, following Eq.~(\ref{lambdanu}), the viscosity scale grows as $\lambda_{\nu}\propto\lambda_H^{1/2}\propto t^{1/2}\propto a^{3/4}$. On the other hand, $\lambda_n\propto a$ always. This means that even wavelengths that are initially smaller than the viscous length will eventually cross outside.

\begin{figure}[tbp]
\begin{center}
\includegraphics[width=\columnwidth]{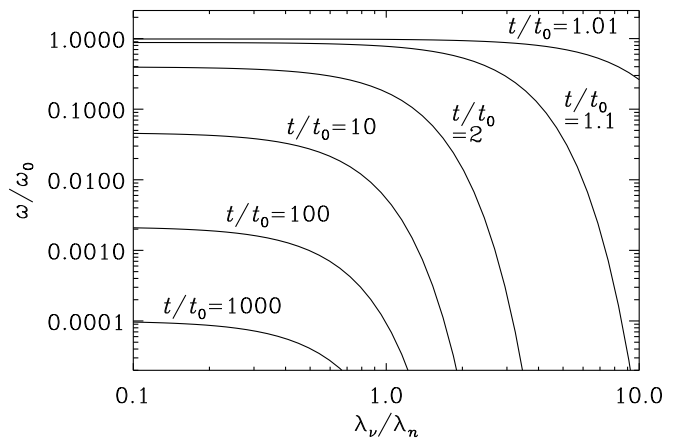}\quad
\end{center}
\caption{Evolution of vorticity perturbations, with respect to the scale of the mode, at different times (see Eq.~(\ref{vomega})). Initially, when $t/t_0\simeq1$, all wavelengths have essentially the same amount of rotation ($\omega/\omega_0\simeq1$ almost everywhere). As time goes by, the vorticity decays away. Dissipation is stronger on small scales (i.e.~for $\lambda_{\nu}/\lambda_n\gg1$), where the viscous effects dominate. On large lengths, where the decay is mainly due to the expansion, the vorticity survives longer.}  \label{fig2}
\end{figure}

The effect of fluid viscosity, with respect to the scale of the rotational distortion at different times, is plotted in Fig.~\ref{fig2}. There, one can see that viscosity dictates the evolution of rotational perturbations that lie deep inside the viscous scale, while much larger lengths are only affected by the background expansion. To be precise, when $\lambda_{\nu}/\lambda_n\gg1$, we find that
\begin{equation}
\omega= \omega_0\exp\left\{-4\left({\lambda_{\nu}\over\lambda_n}\right)_0^2 \left[1-\left({t_0\over t}\right)^{1/3}\right]\right\}\,.  \label{ssvomega}
\end{equation}
Therefore, modes that have remained well inside the viscous scale long enough have
\begin{equation}
\omega= \omega_0 \exp\left[-4\left({\lambda_{\nu}\over\lambda_n}\right)_0^2\right]\,,  \label{ltssvomega}
\end{equation}
when they cross $\lambda_{\nu}$. Afterwards, the background expansion dominates and these modes evolve according to solution (\ref{ltvomega}). In other words, the factor
$\exp\left[-4\left(\lambda_{\nu}/\lambda_n\right)_0^2\right]$ in
Eq.~(\ref{ltvomega}) expresses the total vorticity reduction that is caused by the fluid viscosity on small scales, while the power-law reflects its subsequent decay due to the universal expansion. Clearly, the dissipative effect of viscosity on modes that remained inside the viscous scale for long (i.e.~those with $\lambda_{\nu}/\lambda_n\gg1$ initially) is very strong, leaving no residual vorticity for all practical purposes. Finally, we should also note that, in analogy to Silk-damping~\cite{S}, viscous dissipation could potentially affect scales considerably larger than $\lambda_{\mu}$. Investigating whether and how this happens, however, goes beyond the scope of the present paper.

\section{Magnetized rotation}
\subsection{Magnetically induced vorticity}
Let us now consider the rotational behavior of a self-gravitating Newtonian medium in the presence of a magnetic field. The vorticity evolution is still monitored by Eq.~(\ref{dotomega}), though the shape of the Navier-Stokes formula changes. Assuming zero fluid pressure and very high electrical conductivity (i.e.\ adopting the ideal MHD approximation), we have
\begin{equation}
\rho A_a= -\varepsilon_{abc}B^b{\rm curl}B^c=
-{1\over2}\,\partial_a B^2+ B^b\partial_bB_a\,,  \label{mNS}
\end{equation}
where the right-hand side provides the Lorentz force and we are employing Heaviside-Lorentz units for the Maxwell field.\footnote{One can also obtain (\ref{mNS}) from the Navier-Stokes equation of an imperfect fluid, namely from (\ref{vNS}). Indeed, recalling that the $B$-field corresponds to a imperfect medium with $p=B^2/6$ and $\pi_{ab}= \Pi_{ab}=(B^2/3)\delta_{ab}-B_aB_b$, with $\partial^aB_a=0$ at the ideal-MHD limit, it is straightforward to show that expression (\ref{vNS}) leads to Eq.~(\ref{mNS}).}
In many applications it helps to split the Lorentz force into a pressure and a tension part, with the former reflecting the tendency of the magnetic force-lines to push each other apart and the latter their ``desire'' to remain straight.

Taking the ``curl'' of (\ref{mNS}) and then substituting the resulting expression into the right-hand side of (\ref{dotomega}), while keeping in mind that $\partial^aB_a=0$ at the ideal-MHD limit, we arrive at (see Appendix~\ref{A2} for more details)
\begin{eqnarray}
\dot{\omega}_a = &\!-\!& {2\over3}\,\Theta\omega_a
-{1\over4\rho}\,\varepsilon_{abc}\partial^b
\left(\ln\rho\right)\partial^cB^2 \nonumber\\ &\!+\!& {1\over2\rho}\,\varepsilon_{abc}\partial^b\left(\ln\rho\right)
B_d\partial^{\langle d}B^{c\rangle}- {1\over4\rho}\,
{\rm curl}B_b\partial^b\left(\ln\rho\right)B_a \nonumber\\
&\!+\!&{1\over4\rho}\,B_b\partial^b\left(\ln\rho\right)
{\rm curl}B_a+ {1\over2\rho}\,{\rm curl}B^b\partial_{(b}B_{a)}
\nonumber\\ &\!-\!& {1\over2\rho}\,B^b\partial_b{\rm curl}B_a+
\sigma_{ab}\omega^b\,.  \label{Bdotomega}
\end{eqnarray}
According to the above, which provides an irreducible decomposition of the vorticity propagation formula, the $B$-field will generally act as a source of rotation, even in the case of a homogeneous density distribution.\footnote{As in Eq.~(\ref{vsource}) before, one can replace the gradient $\partial_a(\ln\rho)$ with the vector $\Delta_a$, where the latter defines the (non-dimensionless) density contrast between two neighboring flow lines.} We also note that, of the six magnetic source-terms on the right-hand side of (\ref{Bdotomega}), only the first comes from the pressure component of the Lorentz force; see Eq.~(\ref{mNS}). This term may be seen as the magnetic analogue of the baroclinic stress that emerges in non-barotropic fluids (see~\cite{D-SB} and also compare to Eq.~(\ref{vdotomega}) here). The rest of the magnetic source terms in (\ref{Bdotomega}) are due to the field's tension. All these imply that, in the absence of density inhomogeneities, vorticity is generated solely by the magnetic tension.

\subsection{Magnetized rotating universes}\label{ssMRUs}
Expression (\ref{Bdotomega}) describes the nonlinear rotational behavior of a self-gravitating, Newtonian medium, of zero pressure and high electrical conductivity. When linearized around the (non-magnetized) Newtonian analogue of the Einstein-de Sitter universe, the aforementioned relation reduces to\footnote{When linearizing (\ref{Bdotomega}) we have treated the magnetic pressure (isotropic and anisotropic) as linear perturbation. This ensures that both $B^2$ and $\Pi_{ab}$ are first-order variables, while the magnetic vector and its spatial gradients are 1/2-order perturbations. Then, the magnetic terms on the right-hand side of (\ref{EdSmdotomega}) have perturbative order one.}
\begin{equation}
\dot{\omega}_a= -2H\omega_a+ {1\over2\rho}\,{\rm curl}B^b\partial_{(b}B_{a)}- {1\over2\rho}\,B^b\partial_b{\rm curl}B_a\,.  \label{EdSmdotomega}
\end{equation}
Therefore, large-scale magnetic fields can act as sources of linear vorticity in cosmological environments~\cite{W}. The difference with the nonlinear expression (\ref{Bdotomega}) given before is that now the sole source of vorticity is the magnetic tension. In other words, it is the negative pressure that the $B$-field exerts along its own direction which generates linear rotational perturbations. We should also note that the last term in Eq.~(\ref{EdSmdotomega}) will not lead to vorticity generation when the ``curl'' of the field remains invariant along its own direction. This happens, for example, when the field lines are straight or circular.

Expressions (\ref{Bdotomega}) and (\ref{EdSmdotomega}) also allow one to look at the magnetic effect on pre-existing vorticity. For example, the presence of the field will generally cause precession effects by changing the direction of the vorticity vector, namely the rotational axis of the fluid. Further insight can be obtained by taking the convective derivative of (\ref{EdSmdotomega}). Then, keeping in mind that $\dot{H}=-\kappa\rho/2$ and $\dot{\rho}= -3H\rho$ in the Einstein-de Sitter background -- see Eqs.~(\ref{N1}b) and (\ref{N2}) -- gives
\begin{equation}
\ddot{\omega}_a= -5H\dot{\omega}_a- \kappa\rho\omega_a\,.  \label{EdSmddotomega1}
\end{equation}
Note that in deriving the above we have also used the 1/2-order relations $\dot{B}_a=-2HB_a$, which is the induction equation, and the associated commutation laws $(\partial_bB_a)^{\cdot}= -3H\partial_bB_a$, $({\rm curl}B_a)^{\cdot}=-3H{\rm curl}B_a$ and $(\partial_b{\rm curl}B_a)^{\cdot}=-4H\partial_b{\rm curl}B_a$ (see also in Appendix~\ref{A3} for further details). In addition, to eliminate the magnetic terms, we have re-employed expression (\ref{EdSmdotomega}). Finally, recalling that $H=2/3t$ and $\kappa\rho=4/3t^2$ in an Einstein-de Sitter universe (see~Eq.~(\ref{EdS})), expression (\ref{EdSmddotomega1}) reads
\begin{equation}
\ddot{\omega}_a= -{10\over3}\,t^{-1}\,\dot{\omega}_a- {4\over3}\,t^{-2}\,\omega_a\,.  \label{EdSmddotomega2}
\end{equation}
The latter accepts the solution
\begin{equation}
\omega_a\propto t^{-1}\propto a^{-3/2}\,,  \label{EdSomega}
\end{equation}
since $a\propto t^{2/3}$ to zeroth order. This result shows that magnetized vortices decay inversely proportional to time, as opposed to their $\omega_a\propto t^{-4/3}\propto a^{-2}$ evolution in the non-magnetic case. The presence of the $B$-field has reduced the decay rate of linear vorticity perturbations. In other words, a magnetized Einstein-de Sitter universe should contain more residual rotation than its non-magnetized counterpart. An analogous conclusion has also been reached, through a relativistic analysis, indirectly -- after studying the magnetic effects on rotational (vortex-like) density perturbations in an almost-FRW universe~\cite{TM}.

Unlike viscosity, the linear magnetic effect on vorticity is scale independent. In practice, this means that the associated domain of influence is essentially determined by the coherence length of the $B$-field. Also in contrast to viscosity, the magnetic presence seems to ``favor'' both the generation and the survival of linear vorticity perturbations. The latter effect, in particular, should be directly attributed to the tension properties of the magnetic field lines, namely to the negative pressure the $B$-field exerted along its own direction.\footnote{When dealing with linear density (scalar) perturbations, only the positive magnetic pressure is accounted for. In that case, the effect of the field is to reduce the growth rate of the inhomogeneities, or to increase the associated Jeans scale (i.e.~the region where pressure support prevents these distortions from growing)~\cite{TM}.}

\section{Discussion}
Realistic media do not behave like perfect fluids and magnetic fields, of various scale and strength, appear to be everywhere in the universe. At the same time,  cosmic rotation still poses a number of fundamental questions that remain as yet unanswered. In this respect, it would be interesting to see whether viscosity and magnetism have something new to say about vorticity in general. So far, both agents have been known to act as sources of rotation, though their effects on pre-existing vorticity are not clear yet.

The present article adopts the Newtonian approach to investigate the rotational behavior of a viscous fluid and subsequently that of a magnetized medium. At first, our analysis is fully nonlinear and confirms that both viscosity and magnetism can generate vorticity. We also show that the former of these two agents does so through the coupling of the shear with certain types of density inhomogeneities. Magnetism, on the other hand, does not necessarily need an inhomogeneous matter distribution to source rotational distortions. The $B$-field generates vorticity by means of its Lorentz force, especially via the tension component of the latter.

Assuming the Newtonian analogue of the Einstein-de Sitter universe and then linearizing our equations around this background, we looked into the effects of viscosity and magnetism on rotational cosmological perturbations. We found that the former effects are negative. Viscosity generally leads to vorticity dissipation, though (for non-relativistic species) its role is typically confined to relatively small lengths. Nevertheless, further study is required to establish the actual range of the dissipative effects. Here, we have taken the first step in this direction by identifying the characteristic ``viscous scale'', below which viscosity dominates and beyond which the universal expansion takes over.

In contrast, the magnetic presence seems to favor rotation. More specifically, after linearizing our equations around the aforementioned Einstein-de Sitter background, we found that the $B$-field slows down the standard decay rate of rotational cosmological perturbations. In particular, vorticity no longer obeys the standard $a^{-2}$ decay-law, but drops as $a^{-3/2}$ ($a$ is the cosmological scale factor). Also, the magnetic effects are scale-independent, which means that their domain of influence is decided by the coherence length of the $B$-field. Analogous results were also obtained when studying vortex-like perturbations in the density distribution of a magnetized medium. Overall, it appears as though magnetism favors both the emergence and the survival of cosmic rotation. Put another way, magnetized universes should contain more residual vorticity than their magnetic-free counterparts.

In closing, we should also note that it would be worth-studying the above described effects using a relativistic approach. This will enable one to involve highly energetic particles, look deep into the pre-decoupling universe and also allow for spacetimes more complex than the Einstein-de Sitter universe.

\acknowledgments{CGT wishes to thank Nordita for the hospitality during his visit, when the main part of the study took place. This work was partly supported by the European Research Council (AstroDyn Research Project No.~227952).}

\appendix

\section{The viscous source term}\label{A1}
Viscosity generates vorticity through the third term in the right-hand side of Eq.~(\ref{vdotomega}). Further insight can be obtained by decomposing this source term to its irreducible components. To begin with, we have
\begin{eqnarray}
\varepsilon_{abc}\partial^b \left[\sigma^{cd}\partial_d\left(\ln\rho\right)\right]&=& \varepsilon_{abc}\partial^b\sigma^{cd}\partial_d\left(\ln\rho\right) \nonumber\\ &&+ \varepsilon_{abc}\sigma^{cd}\partial^b\partial_d\left(\ln\rho\right) \nonumber\\ &=&\varepsilon_{abc}\partial^b\sigma^{cd}\Delta_d \nonumber\\ &&+\varepsilon_{abc}\sigma^c{}_d\Sigma^{db}\,,  \label{App1}
\end{eqnarray}
where $\Delta_a=\partial_a(\ln\rho)$ and $\Sigma_{ab}=\partial_{\langle b}\partial_{a\rangle}(\ln\rho)$ by definition. The antisymmetric part of $\partial_b\partial_a\left(\ln\rho\right)$ is identically zero, while its trace has no contribution to decomposition (\ref{App1}), due to the total antisymmetry of the Levi-Civita tensor and the symmetry properties of the shear. Moreover, recalling that ${\rm curl}\sigma_{ab}=\varepsilon_{cd\langle a}\partial^c\sigma^d{}_{b\rangle}$ and the trace-free nature of the shear tensor, we may write
\begin{equation}
{\rm curl}\sigma_{ab}= {1\over2}\,\varepsilon_{cda}\partial^c\sigma^d{}_b+ {1\over2}\,\varepsilon_{cdb}\partial^c\sigma^d{}_a\,.  \label{App2}
\end{equation}
Also, given that ${\rm curl}\sigma_{ab}=-\partial_{\langle b}\omega_{a\rangle}$ (see constraint (\ref{con1}b)), the above recasts into the auxiliary relation
\begin{equation}
\Delta^b\varepsilon_{cda}\partial^c\sigma^d{}_b= -2\Delta^b\partial_{\langle b}\omega_{a\rangle}- \Delta^b\varepsilon_{bcd}\partial^c\sigma^d{}_a\,.  \label{App3}
\end{equation}
Finally, substituting this result into the right-hand side of (\ref{App1}) and employing some straightforward tensor algebra, one arrives at decomposition (\ref{vsource}). The latter shows that only part of the viscous source term in Eq.~(\ref{vdotomega}) can actually generate vorticity.

\section{Nonlinear magnetized vorticity}\label{A2}
The rotational evolution of a medium is governed by Eq.~(\ref{dotomega}), where the key term is the ``curl'' of the acceleration vector. By definition, the latter is ${\rm curl}A_a= \varepsilon_{abc}\partial^bA^c$. Then, starting from the magnetized Navier-Stokes equation (see expression (\ref{mNS})), we first arrive at
\begin{eqnarray}
{\rm curl}A_a&=& {1\over2\rho}\,\varepsilon_{abc}\partial^b \left(\ln\rho\right)\partial^cB^2- {1\over\rho}\,\varepsilon_{abc}\partial^b\left(\ln\rho\right) B_d\partial^dB^c \nonumber\\ &&+ {1\over\rho}\,\varepsilon_{ab}{}^c\partial^bB^d\partial_dB_c+ {1\over\rho}\,B^b\partial_b{\rm curl}B_a\,.  \label{App4}
\end{eqnarray}
Recalling that $\partial^aB_a=0$ at the ideal-MHD limit and using the total antisymmetry of the Levi-Civita tensor, the third term on the right-hand side of the above gives
\begin{eqnarray}
\varepsilon_{ab}{}^c\partial^bB^d\partial_dB_c&=& 2\varepsilon_{ab}{}^c\partial^{(b}B^{d)}\partial_{[d}B_{c]} \nonumber\\&=& -{\rm curl}B^b\partial_{(b}B_{a)}\,,  \label{App5}
\end{eqnarray}
since that $2\partial_{[a}B_{b]}=\varepsilon_{abc}{\rm curl}B^c$. In addition, given that $\partial_bB_a=\partial_{\langle b}B_{a\rangle}+\partial_{[b}B_{a]}$, the second term in the right-hand side of (\ref{App4}) decomposes to
\begin{eqnarray}
\varepsilon_{abc}\partial^b\left(\ln\rho\right) B_d\partial^dB^c&=& \varepsilon_{abc}\partial^b\left(\ln\rho\right) B_d\partial^{\langle d}B^{c\rangle} \nonumber\\ &&+ {1\over2}\,B_b\partial^b\left(\ln\rho\right){\rm curl}B_a \nonumber\\ &&-{1\over2}\,{\rm curl}B_b \partial^b\left(\ln\rho\right)B_a\,.  \label{App6}
\end{eqnarray}
Substituting the last two auxiliary relations back into Eq.~(\ref{App4}) and the resulting expression into (\ref{dotomega}) we obtain Eq.~(\ref{Bdotomega}). The latter monitors the nonlinear evolution of rotational perturbations in a magnetized, highly conductive medium.

\section{Linear commutation laws}\label{A3}
In Euclidean spaces, simple partial derivatives (temporal or spatial) commute. This is not the case, however, when convective derivatives are involved. Then, one needs to employ the associated commutation laws. In \S~\ref{ssMRUs}, we were only interested in the linear version of these formulae. Hence, starting from the convective derivative of $\partial_bB_a$, using decomposition (\ref{pbva}) and then linearizing around an Einstein-de Sitter universe, one finds
\begin{eqnarray}
\left(\partial_bB_a\right)^{\cdot}&=& \partial_t\partial_bB_a+ v^c\partial_c\partial_bB_a \nonumber\\ &=&\partial_b\partial_tB_a+ \partial_b\left(v^c\partial_cB_a\right)- \partial_bv^c\partial_cB_a \nonumber\\ &=&\partial_b\dot{B}_a- \left({1\over3}\, \Theta\delta^c{}_b+\sigma^c{}_b +\varepsilon^c{}_{bd}\omega^d\right)\partial_cB_a \nonumber\\ &=&-3H\partial_bB_a\,.  \label{App7}
\end{eqnarray}
Also note that $\dot{B}_a=-2HB_a$ to first order, while $\Theta=3H$ and $\sigma_{ab}=0=\omega_a$ in the unperturbed background. Similarly, we have
\begin{eqnarray}
\left({\rm curl}B_a\right)^{\cdot}&=& \varepsilon_{abc}\partial_t\partial^bB^c+ \varepsilon_{abc}v^d\partial_d\partial^bB^c \nonumber\\ &=&\varepsilon_{abc}\partial^b\partial_tB^c+ \varepsilon_{abc}\partial^b\left(v^d\partial_dB^c\right) \nonumber\\ &&-\varepsilon_{abc}\partial^bv^d\partial_dB^c \nonumber\\ &=&\varepsilon_{abc}\partial^b\dot{B}^c- H{\rm curl}B_a \nonumber\\ &=&-3H{\rm curl}B_a\,,  \label{App8}
\end{eqnarray}
to linear order. Finally, proceeding in an analogous way and then combining the last two results, one arrives at
\begin{equation}
\left(\partial_b{\rm curl}B_a\right)^{\cdot}= -4H\partial_b{\rm curl}B_a\,.  \label{App9}
\end{equation}
On using these commutation laws, between convective and spatial partial derivatives, one can begin from Eq.~(\ref{EdSmdotomega}) and obtain expression (\ref{EdSmddotomega1}). The latter monitors the linear evolution of magnetized rotational perturbations on an Newtonian Einstein-de Sitter universe.

\end{document}